\documentclass[12pt]{article}
\usepackage{latexsym,amsmath,amssymb,graphicx}
\renewcommand{\thefootnote}{\fnsymbol{footnote}}

\usepackage{amsmath}
\usepackage{amsfonts}

\numberwithin{equation}{section}

\def\doubleset#1#2{\bgroup%
\def\doit#1#2{%
\setbox\dblsetbox=\hbox{$\cstyle #1$}%
\raise#2\ht\dblsetbox\copy\dblsetbox%
\hskip-\wd\dblsetbox%
\raise-#2\ht\dblsetbox\box\dblsetbox}%
\mathchoice%
{\def\cstyle{\displaystyle}\doit#1#2}%
{\def\cstyle{\textstyle}\doit#1#2}%
{\def\cstyle{\scriptstyle}\doit#1#2}%
{\def\cstyle{\scriptscriptstyle}\doit#1#2}\egroup}
\def\underarrow#1{\vbox{\ialign{##\crcr$\hfil\displaystyle
 {#1}\hfil$\crcr\noalign{\kern1pt\nointerlineskip}$\longrightarrow$\crcr}}}

\newbox\dblsetbox

\newlength{\extraspace}
\newlength{\extraspaces}
\newcommand{\be}{\begin{equation}
\addtolength{\abovedisplayskip}{\extraspaces}
\addtolength{\belowdisplayskip}{\extraspaces}
\addtolength{\abovedisplayshortskip}{\extraspace}
\addtolength{\belowdisplayshortskip}{\extraspace}}
\newcommand{\ee}{\end{equation}}

\newcommand{\ba}{\begin{eqnarray}
\addtolength{\abovedisplayskip}{\extraspaces}
\addtolength{\belowdisplayskip}{\extraspaces}
\addtolength{\abovedisplayshortskip}{\extraspace}
\addtolength{\belowdisplayshortskip}{\extraspace}}
\newcommand{\ea}{\end{eqnarray}}

\newcommand{\bd}{\begin{displaymath}
\addtolength{\abovedisplayskip}{\extraspaces}
\addtolength{\belowdisplayskip}{\extraspaces}
\addtolength{\abovedisplayshortskip}{\extraspace}
\addtolength{\belowdisplayshortskip}{\extraspace}}
\newcommand{\ed}{\end{displaymath}}

\newcounter{saveeqn}

\newcommand{\newsection}[1]{
\vspace{12mm} \pagebreak[3] \addtocounter{section}{1}
\setcounter{equation}{0} \setcounter{subsection}{0}
\noindent{\bf \thesection. #1} \nopagebreak
\medskip
\nopagebreak
\addcontentsline{toc}{section}{\thesection. #1}}

\newcommand{\newsubsection}[1]{
\vspace{0.8cm} \pagebreak[3] \addtocounter{subsection}{1}
\setcounter{subsubsection}{0}
\noindent{ \it \thesubsection. #1} \nopagebreak \vspace{2mm}
\nopagebreak
\addcontentsline{toc}{subsection}{\thesubsection. #1}}

\begin{document}
\addtolength{\baselineskip}{1.5mm}

\thispagestyle{empty}

\vbox{} \vspace{2.0cm}

\begin{center}
\centerline{\LARGE{Surface Operators in Abelian Gauge Theory}}
\bigskip

\vspace{2.0cm}

{Meng-Chwan~Tan \footnote{On leave of absence from the National
University of Singapore.}}
\\[2mm]
{\it California Institute of Technology, \\
Pasadena, CA 91125, USA} \\ [1mm]
e-mail: mengchwan@theory.caltech.edu\\
\end{center}

\vspace{2.0 cm}

\centerline{\bf Abstract}\smallskip \noindent

We consider arbitrary embeddings of surface operators in a pure, non-supersymmetric abelian gauge theory on spin (or non-spin) four-manifolds. For any surface operator with a priori $\it{simultaneously}$  $\text{\it{non-vanishing}}$ parameters, we explicitly show that the parameters transform naturally under an $SL(2,\mathbb Z)$ (or $\Gamma_0(2)$) duality of the theory. However, for non-trivially-embedded surface operators, $\it exact$ $S$-duality holds only if the quantum parameter effectively vanishes, while the overall $SL(2, \mathbb Z)$ (or $\Gamma_0(2)$) duality holds up to a $c$-number at most, regardless. Via the formalism of duality walls, we furnish an alternative derivation of the transformation of parameters - found also to be consistent with a switch from Wilson to 't Hooft loop operators under $S$-duality. With any background embedding of surface operators, the partition function and the correlation functions of non-singular, gauge-invariant local operators on any $\it{curved}$ four-manifold, are found to transform like modular forms under the respective duality groups.

 \newpage

\renewcommand{\thefootnote}{\arabic{footnote}}
\setcounter{footnote}{0}

\tableofcontents


\vspace{0.0cm}

\newsection{Introduction and Summary}

Surface operators are higher-dimensional analogues of the usual Wilson and 't Hooft loop operators in gauge theory, that are supported on a codimension two submanifold of spacetime. They are defined by specifying a certain type of singularity in the relevant fields as one approaches the submanifold. Such operators were first used to probe the dynamics of gauge theory and black holes in~\cite{Preskill}-\cite{Bucher}. Thereafter, they appeared in the mathematical literature in an application to Donaldson theory~\cite{Mrowka}, and in the relation between instantons, Seiberg-Witten theory and integrable systems~\cite{Braverman, Etingof}.

More recently, in an effort to furnish a gauge-theoretic interpretation of the geometric Langlands program with ramification, surface operators have also been considered in a twisted version of ${\cal N} = 4$ supersymmetric Yang-Mills theory in four dimensions~\cite{Gukov-Witten}. They have also made an appearance in the context of the AdS/CFT correspondence between ${\cal N}=4$ SYM and type IIB supergravity~\cite{2}-\cite{8}, whereby the proposed action of the $SL(2,\mathbb Z)$ duality group on the parameters of a surface operator found in~\cite{Gukov-Witten}, has been shown to be consistent in a dual type IIB supergravity description in~\cite{8}.

To date, there has not been an explicit way to prove that the parameters of a surface operator in the ${\cal N}=4$ gauge theory ought to transform as proposed in~\cite{Gukov-Witten}. Moreover, most examples involve only trivial embeddings of surface operators in spacetime; not much is known about the action of the $SL(2, \mathbb Z)$ duality group on the parameters of surface operators that are non-trivially embedded. Nevertheless, one can do much better with a simpler theory that also possesses an $SL(2,\mathbb Z)$ duality; namely, a pure, non-supersymmetric abelian gauge theory. In fact, it has been explicitly demonstrated in~\cite{Gukov-Witten} using a pure $U(1)$ gauge theory, that for a surface operator with one of its two parameters set to zero, the remaining parameter does indeed transform as proposed under $S$-duality. One can therefore hope that a rigorous understanding of surface operators in this non-supersymmetric, abelian setting, would help shed some light on the above issues.

\smallskip\noindent{\it A Summary of the Paper}

In this paper, we shall consider arbitrary embeddings of surface operators in the pure $U(1)$ gauge theory on spin (or non-spin) four-manifolds. We shall derive explicitly, the transformation of the surface operator parameters - that are a priori $\it{simultaneously}$ $\it{non}$-$\it{vanishing}$ - under an $SL(2,\mathbb Z)$ (or a congruence subgroup $\Gamma_0(2)$) duality of the gauge theory. We find an agreement with the proposal put forth in~\cite{Gukov-Witten}, except when a surface operator is non-trivially-embedded, in which case the relevant dualities hold under certain conditions only. By considering a simple  correlation function between a Wilson or 't Hooft loop operator and a surface operator, we find that the transformation of parameters is consistent with a switch from Wilson to 't Hooft loop operators  under $S$-duality. Via the formalism of duality walls, we shall also provide an alternative derivation of the transformation of parameters. Last but not least, we analyse the partition function and the correlation functions of non-singular, gauge-invariant local operators in the presence of an arbitrarily-embedded surface operator in the background. We find that the partition function and the correlation functions all behave like modular forms of $SL(2,\mathbb Z)$ (or $\Gamma_0(2)$), albeit with different modular weights.



\newsection{Surface Operators in Pure $U(1)$ Gauge Theory}

\vspace{-0.0cm}

\newsubsection{Description of the Relevant Surface Operators}

\smallskip\noindent{\it The Parameters $\alpha$ and $\eta$}

In this paper, we shall consider surface operators that are supported on an arbitrary two-submanifold $D$ in pure $U(1)$ gauge theory on a general four-manifold $M$, where $D$ and $M$ are assumed to be oriented.
The surface operator is to be characterised by a gauge field solution that gives rise to a singular field strength as one approaches $D$. In addition, the gauge field solution must be invariant under rotations of the plane $D'$ normal to $D$.

An example of such a gauge field solution is
\be
A = \alpha d\theta,
\label{A}
\ee
where $\alpha$ is a parameter valued in $U(1)$,\footnote{Such a parameter of the gauge field ought to be valued in the (real) Lie algebra ${\frak u}(1)$. However, as explained in \cite{Gukov-Witten}, one can shift the parameter $\alpha \to \alpha + u$ in a particular gauge transformation, whereby $\text{exp}(2\pi u) =1$. The only invariant of such a gauge transformation is the monodromy $\text{exp}(-2\pi \alpha)$ of the gauge field $A$ around a circle of constant $r$. Hence, $\alpha$ must take values in $\mathbb R / \mathbb Z$ instead.} and $\theta$ is the angular component of the coordinate $z = r e^{i\theta}$ on $D'$. Noting that $d(d\theta) = 2 \pi \delta_D$ (where $\delta_D$ is a two-form delta function supported at the origin of $z$, that is also Poincar\'e dual to $D$), we find the corresponding field strength to be
\be
F = 2 \pi \alpha \delta_D.
\label{F}
\ee
As required, $F$ is singular as one approaches $D$.

However, note from footnote 1 that we are free to shift $\alpha$ by $u$ via a gauge transformation. As such, this definition of $F$ appears to be unnatural. This can be remedied by lifting $\alpha$ in (\ref{F}) from $U(1)$ to ${\frak u}(1)$, such that it is no longer true that $\alpha \sim \alpha + u$, that is, $F$, when restricted to $D$, is ${\frak u}(1)$-valued. Equivalently, this corresponds to finding an extension of the $U(1)$-bundle $E$ on $M$ with connection $A$, over $D$ (whereby due to the singularity along $D$, $E$ is originally defined on the complement of $D$ in $M$ only). Such an extension exists whenever $E$ is a $U(1)$-bundle on $M$. Thus, the definition of $F$ in (\ref{F}) actually makes sense.

Notice that since we have an extension of the bundle $E$ over $D$, we roughly have an abelian gauge theory in two dimensions on $D$. As such, one can introduce a two-dimensional theta-like angle $\eta$ as an additional quantum parameter, which enters in the path integral via the phase
\be
\textrm{exp}\left(-i\eta \int_D F\right).
\label{eta term}
\ee
Notice that $\eta$ must therefore take values in $\mathbb R / \mathbb Z$, since the integrated first Chern class $\int_D F/2 \pi$ of the $U(1)$-bundle $E \to D$, is an integer. Just like $\alpha$, one can shift $\eta$ (by an integral lattice) whilst leaving the theory invariant.

\vspace{0.3cm}

\smallskip\noindent{\it A Point on Non-Trivially-Embedded Surface Operators}

  More can also be said about the parameter $\alpha$ as follows. In the case when the surface operator is trivially-embedded in $M$, that is, $M = D' \times D$ and the normal bundle to $D$ is hence trivial, the self intersection number
\be
D \cap D = \int_{M} \delta_D \wedge \delta_D
\label{DcapD}
\ee
vanishes. On the other hand, for a non-trivially-embedded surface operator supported on $D \subset M$, the normal bundle is non-trivial, and the intersection number is non-zero. The surface operator is then defined by the gauge field with singularity in (\ref{A}) in each normal plane.

When the intersection number is non-zero, or rather for non-trivially embedded surface operators, there is a restriction on the values that $\alpha$ can take. To explain this, first note that since $F = 2 \pi \alpha \delta_D$ near $D$, we find, using (\ref{DcapD}), that $\int_D F/2 \pi  = \alpha \ D \cap D \ \textrm{mod} \ \mathbb Z$. Since the integrated first Chern class $\int_D F/2 \pi$ is always an integer, we must have
\be
\alpha \ D \cap D \in \mathbb Z.
\label{intersection number}
\ee
In particular, the only gauge transformations that can be defined globally along $D$, are those that shift $\alpha$ in such a way as to maintain the condition (\ref{intersection number}). This point will be important later.

\newsubsection{Action of Duality on Trivially-Embedded Surface Operators}

\smallskip\noindent{\it Action of $S$-duality}

We shall now discuss the case of a trivially-embedded surface operator - with simultaneously non-vanishing parameters $(\alpha, \eta)$ - in the pure $U(1)$ gauge theory on $\it{any}$ closed four-manifold $M$. Our first objective is to prove explicitly that the parameters transform as
\be
(\alpha, \eta) \to (\eta, - \alpha)
\label{parameter transform 1}
\ee
under the $S$-duality transformation $S : \tau \to - 1/ \tau$ of the gauge theory. Here, $\tau = \theta/2\pi + 4\pi i /g^2$ is the complexified gauge coupling parameter. To this end, we shall adopt the approach taken in~\cite{abelian S-duality}.

Before we proceed further, we would like to mention again that for a particular surface operator with parameters $(\alpha, \eta) = (0, \eta)$, the above claim has already been explicitly proven in $\S$2.4 of~\cite{Gukov-Witten}. However, it is felt that the proof in~\cite{Gukov-Witten} could have been made more concrete via a less ad-hoc conclusion. We hope to remedy this with our proof involving a general surface operator with non-vanishing $\eta$ $\it{and}$ $\alpha$.

Getting back to our discussion, note that in the pure $U(1)$ gauge theory, we have a gauge field $A$ (which is locally a real one-form, since $\frak u(1)$ is real) that we can think of as a connection on a principle $U(1)$-bundle $\cal L$ on $M$, with curvature $F = dA$. We shall take the action to be (in Euclidean signature)
\begin{eqnarray}
\label{I_1}
\bf{I} & = & {1\over 8 \pi} \int_M d^4x {\sqrt h} \left( {{4\pi} \over g^2} F_{mn}F^{mn} - {{i \theta}\over 2 \pi}{1 \over 2} \epsilon_{mnpq} F^{mn}F^{pq} \right) \nonumber \\
& = & {1\over g^2} \int_M F \wedge \star F - {{i \theta}\over 8 \pi^2}\int_M F \wedge F,
\end{eqnarray}
where $h$ is the metric on $M$, $\epsilon_{mnpq}$ is the Levi-Civita antisymmetric tensor, and the Hodge-star operator acts on any two-form in $M$ as $\star (dx^m \wedge dx^n) = {1\over 2} \epsilon_{mnpq} dx^p \wedge dx^q$. Noting that $F^{\pm} = {1\over 2} (F \pm \star F)$ are the self-dual and anti-self-dual projections of $F$, we can alternatively write the action as
\begin{eqnarray}
\label{I_2}
\bf{I}_{\tau} & = & -{i \over 8\pi} \int_M d^4x \sqrt h \left(\tau F_{mn}^+ F^{+ mn}  - \bar\tau F_{mn}^- F^{- mn} \right) \nonumber \\
& = & -{{i \tau} \over 4\pi} \int_M F^+ \wedge \star F^+  + {{i \bar\tau} \over 4\pi} \int_M F^- \wedge \star F^-.
\end{eqnarray}

Note that on any closed four-manifold $M$, $c_1({\cal L})^2 = \int_M (F/ 2\pi) \wedge (F / 2 \pi)$ is always an integer, where $c_1({\cal L})$ is the first Chern class of $\cal L$. Since the action $\bf{I}$ appears in the quantum theory through the factor $\textrm{exp}(-\bf{I})$ in the path integral (that is, the quantum theory is unaffected when $\bf{I}$ is shifted by $2 \pi i \mathbb Z$), the quantum theory is invariant under $\theta \to \theta + 4 \pi$ or $\tau \to \tau +2$. However, if $M$ is a closed spin manifold, $c_1({\cal L})^2$ is always an even integer. Then, the quantum theory will be invariant under $\theta \to \theta + 2\pi$ or $\tau \to \tau +1$. Together with the invariance of $\bf{I}$ under the $S$-duality transformation $\tau \to -{1 \over \tau}$ (see \cite{abelian S-duality} for an explicit proof of this statement), we find that one can at least have full modular invariance (that is, invariance under the full $SL(2,\mathbb Z)$ group generated by $S:\tau \to -{1 \over \tau}$ and $T:\tau \to \tau +1$) only when $M$ is spin. On the other hand, if $M$ is non-spin, the theory would be invariant under the transformations $S$ and $ST^2S$ which generate the congruence subgroup $\Gamma_0(2)$ of $SL(2,\mathbb Z)$.

When one introduces a surface operator into the theory, certain modifications need to be made to the above description. Firstly, recall that the presence of a surface operator results in a singularity of the field strength $F$ along its support $D$ (see (\ref{F})). Since the action $\bf{I}_{\tau}$ is quadratic in $F^+$ and $F^-$ with a positive-definite real part, it is potentially divergent. Therefore, in computing the path integral, where one must sum over all inequivalent principle $U(1)$-bundles on $M$, the corresponding connections that will contribute to the computation must then have non-singular curvatures $F' = F - 2\pi \alpha \delta_D$. As such, one might just as well replace $F$ by $F'$ in the above equations, and study instead the action
\begin{eqnarray}
\label{I'tau}
{\bf{I}}'_{\tau} (A) & = & -{{i \tau} \over 4\pi} \int_M {F'}^{+} \wedge \star {F'}^{+} + {{i \bar\tau} \over 4\pi} \int_M {F'}^{-} \wedge \star {F'}^{-} \nonumber \\
& = & -{{i \tau} \over 4\pi} \int_M (F^{+} - 2\pi \alpha \delta^{+}_D) \wedge \star (F^{+}- 2\pi \alpha \delta^{+}_D) + {{i \bar \tau} \over 4\pi} \int_M (F^{-} - 2\pi \alpha \delta^{-}_D) \wedge \star (F^{-} - 2\pi \alpha \delta^{-}_D). \nonumber \\
\end{eqnarray}

As mentioned earlier, one must also include in the action, the theta-like term
\be
{\bf{I}}_{\eta} (A) = i \eta \int_D F'.
\label{I_eta}
\ee
Notice that we have again replaced $F$ with $F'$ in the above term. This is because the singularity in $F$ will result in a highly oscillatory contribution to the path integral that is tantamount to taking its classical limit - an approximation that we do not wish to consider.

We shall now introduce a two-form $\bf{g}$ that is invariant under the usual Maxwell abelian gauge symmetry $A \to A -d\epsilon$ (where $\epsilon$ is a zero-form). We would then like to define the following extended gauge symmetry
\begin{eqnarray}
\label{extended gauge symmetry tx}
A &\to & A + b \nonumber \\
{\bf g} & \to & {\bf g} + db,
\end{eqnarray}
where $b$ is a connection one-form on a principle $U(1)$-bundle $\cal T$ with curvature $db$. If $\cal T$ has trivial curvature with $b = -d\epsilon$, one gets back the usual Maxwell abelian gauge symmetry. Since $A$ is a connection on the bundle $\cal L$, it will mean that $A + b$ is a connection on the bundle $\cal L \otimes \cal T$. For trivial (or flat) $\cal T$, where one just has an ordinary Maxwell theory, it is clear that it suffices to consider only some $\cal L$ in order to define the theory properly. However, in order to generalise the theory to non-trivial $\cal T$ - that is, for $(A +b)$ and $({\bf{g}} + db)$ to be physically valid as a gauge field and two-form, respectively - one must necessarily sum over all $\cal L$'s.

A relevant consequence of an invariance of any theory under (\ref{extended gauge symmetry tx}), is that one is free to shift the periods of $\bf g$ - that is, the integrals of $\bf {g}$ over closed two-dimensional cycles $S \subset M$ - by integer multiples of $2 \pi$:
\be
\int_S {\bf{g}} \to \int_S {\bf{g}} + 2 \pi n, \quad \forall n \in \mathbb Z.
\label{g period}
\ee
(Here, we have made use of the fact that since $db$ is a curvature of a line bundle $\cal T$, we have $\int_S db  = 2\pi \int_S c_1({\cal T}) \in 2\pi \mathbb Z$).

One way to modify the total action $\bf{I}'_{\tau} + \bf{I}_{\eta}$ so that we can have invariance under the transformations (\ref{extended gauge symmetry tx}), is to replace $F'$ with ${\cal F}' = F' - \bf{g}$. However, notice that the resulting theory is trivial and not equivalent to the original theory, because one cannot set $\bf g$ to zero even if we let $b = - d\epsilon$. Nevertheless, one can introduce another abelian gauge field $\bf{w}$, that is a connection one-form on a principle $U(1)$-bundle $\widetilde {\cal L}$ with curvature ${\bf{W}} = d\bf{w}$, and add to the action the term
\be
{\widetilde {\bf{I}}} = {i \over 8 \pi} \int_M d^4 x \sqrt{h} \epsilon^{mnpq}{\bf W}_{mn} {\bf g}_{pq} = {i \over 2 \pi} \int_M {\bf W} \wedge {\bf g}.
\ee
Like any curvature of a line bundle, we have the condition $\int_S {\bf W}/ 2 \pi \in \mathbb Z$. Thus, we find that ${\widetilde {\bf{I}}}$ is invariant mod $2 \pi i \mathbb Z$ under the extended gauge transformation (\ref{extended gauge symmetry tx}). It is also invariant under the gauge transformation ${\bf w} \to {\bf w} - d {\widetilde \epsilon}$, where $\widetilde \epsilon$ is a zero-form on $M$.

Let us now define an extended theory in the fields $(A, {\bf g}, {\bf w})$ with action
\be
{\widehat {\bf I}} (A, {\bf g}, {\bf w}) = {i \over 2 \pi} \int_M {\bf W} \wedge {\bf g} - {{i \tau} \over 4\pi} \int_M {\cal F'}^{+} \wedge \star {\cal F'}^{+} + {{i \bar\tau} \over 4\pi} \int_M {\cal F'}^{-} \wedge \star {\cal F'}^{-} +i \eta \int_D {\cal F}'.
\label{extended action}
\ee
Since under (\ref{extended gauge symmetry tx}), ${\cal F}'$ is manifestly invariant while $\widetilde {\bf I}$ is invariant mod $2 \pi i \mathbb Z$, we find that ${\widehat {\bf I}} (A, {\bf g}, {\bf w})$ will be invariant mod $2 \pi i \mathbb Z$ under (\ref{extended gauge symmetry tx}), as required. It is also invariant under gauge transformations of ${\bf w}$.

We would now like to show that the extended theory with action ${\widehat {\bf I}}(A, {\bf g}, {\bf w})$ is equivalent to the original theory with action $\bf{I}'_{\tau} + \bf{I}_{\eta}$ that we started with. To this end, first note that the (unregularised) partition function of the extended theory can be written as
\be
{1 \over {\textrm{vol}(\cal G)}}{1 \over {\textrm{vol}(\widehat{\cal G})}}{1 \over {\textrm{vol}(\widetilde {\cal G})}} \sum_{{\cal L},{\widetilde {\cal L}}} \int {\cal D}A \ {\cal D}{\bf g} \ {\cal D} {\bf w} \ \textrm{exp} \left(-{\widehat {\bf I}} (A, {\bf g}, {\bf w})\right),
\label{partition function}
\ee
where $\cal G$ and $\widetilde{\cal G}$ denote the group of gauge transformations associated to $A$ and $\bf{w}$, and $\widehat{\cal G}$ denotes the group of extended gauge transformations associated to $\bf g$. Next, let us try to compute the path integral over the $\bf{w}$ fields. To do this, first write  ${\bf w} = {\bf w}_0 + {\bf w}'$, where ${\bf w}_0$ is a fixed connection on the line bundle $\widetilde {\cal L}$. Then, the path integral over the $\bf w$ fields can be written as
\be
{1 \over {\textrm{vol}(\widetilde {\cal G})}} \sum_{\widetilde {\cal L}} \int {\cal D}{\bf w}'  \ \textrm{exp}\left(-  {i \over 2 \pi} \int_M {\bf w}' \wedge d{\bf g}\right) \cdot \textrm{exp}\left( - {i \over 2 \pi} \int_M {\bf W}_0 \wedge {\bf g}\right),
\label{path integral of w}
\ee
where ${\bf W}_0 = d {\bf w}_0$ corresponds to the curvature of the fixed connection ${\bf w}_0$. It is a closed two-form on $M$ in the cohomology $H^2(M)$, as ${\bf w}_0$ is only defined locally as a one-form. Noting that
\be
{1 \over {\textrm{vol}(\widetilde {\cal G})}} \int {\cal D}{\bf w}'  \ \textrm{exp}\left( - {i \over 2 \pi} \int_M {\bf w}' \wedge d{\bf g}\right) = \delta (d {\bf g}),
\ee
one can compute (\ref{path integral of w}) as
\be
 \sum_{{\bf W}_0 \in H^2(M)} \textrm{exp}\left( - i \int_M {\bf W}_0 \wedge {{\bf g} \over 2 \pi}\right) \cdot \delta (d {\bf g}) = \delta \left( \left[{{\bf g} \over 2 \pi}\right] \in \mathbb Z \right) \cdot \delta (d {\bf g}).
\ee
In other words, we have the condition $d {\bf g} = 0$. We also have the condition that $\left[{\bf{g} \over 2 \pi}\right]$ belongs to an integral class, that is, the periods $\int_S \bf{g}$ must take values in $2 \pi \mathbb Z$. The first condition says that one can pick $\bf{g}$ to be a constant two-form. Together with the second condition and (\ref{g period}), one can indeed obtain ${\bf g} = 0$ via the extended gauge transformation (\ref{extended gauge symmetry tx}). By setting ${\bf g} = 0$,  the action ${\widehat {\bf I}}$ reduces to the original action $\bf{I}'_{\tau} + \bf{I}_{\eta}$. Hence, the theory with action ${\widehat {\bf I}}(A, {\bf g}, {\bf w})$ is indeed equivalent to the original theory that we started with.

Now, let us analyse ${\widehat {\bf I}}(A, {\bf g}, {\bf w})$ in a different gauge, namely, one in which we set $A = 0$ via the extended gauge symmetry (\ref{extended gauge symmetry tx}).\footnote{Note that one can only set $A=0$ (that is, to pure gauge) over all of $M$ via the usual gauge transformation $A \to A - d \epsilon$, for $M$ a simply-connected four-manifold. Nevertheless, one can always use the extended gauge transformation of (\ref{extended gauge symmetry tx}) to set $A=0$ for any $M$.} Noting that
\be
\int_M {\bf W} \wedge {\bf g} = \int_M \left ( {\bf W}^+ \wedge \star {\bf g}^+ - {\bf W}^- \wedge \star {\bf g}^-      \right) = \int_M ({\bf W}^+ \cdot {\bf g}^+) - ({\bf W}^- \cdot {\bf g}^-),
\ee
one can write the action in this gauge as
\begin{eqnarray}
{\widehat {\bf I}} ({\bf g}, {\bf w}) & = & {i \over 2 \pi} \int_M ({\bf W}^+ - 2 \pi  \eta \delta^+_D) \cdot {\bf g}^+ - ({\bf W}^- - 2 \pi  \eta \delta^-_D) \cdot {\bf g}^- - {{i \tau} \over 4\pi} \int_M |2 \pi \alpha \delta^+_D + {\bf g}^{+}|^2 \nonumber \\
&& + {{i \bar\tau} \over 4\pi} \int_M |2 \pi \alpha \delta^-_D + {\bf g}^{-}|^2,
\label{2.20}
\end{eqnarray}
where $|k|^2 = k \wedge \star k$ for any two-form $k$. Note that in the above, we have also used the fact that the term $- 2 \pi i \eta \alpha \int_M \delta_D \wedge \delta_D$ - which generically appears in the action - can be set to zero for a trivially-embedded surface operator. If we define
\be
{\bf g}' = {\bf g} + 2 \pi \alpha \delta_D - {1 \over {\tau}} \left ({\bf W}^+ - 2 \pi \eta \delta^+_D \right) -  {1 \over {\bar\tau}} \left ( {\bf W}^- - 2 \pi \eta \delta^-_D \right),
\label{g'}
\ee
we can rewrite the action as
\begin{eqnarray}
\label{I g'}
{\widehat {\bf I}} ({\bf g}', {\bf w}) & = & -{{i \tau} \over {4 \pi}} \int_M |{\bf g}^{'+}|^2 + {{i \bar\tau} \over {4 \pi}} \int_M |{\bf g}^{'-}|^2  + {i \over {4 \pi \tau}} \int_M |{\bf W}^+ - 2 \pi \eta \delta^+_D|^2 - {i \over {4 \pi \bar\tau}} \int_M |{\bf W}^- - 2 \pi \eta \delta^-_D|^2 \nonumber \\
&& - i \alpha \int_D ({\bf W} - 2 \pi \eta \delta_D).
\end{eqnarray}
Then, by integrating out the ${\bf g}^{'+}$ and ${\bf g}^{'-}$ fields classically using the Euler-Lagrange equations, we have
\begin{eqnarray}
{\widehat {\bf I}} ({\bf w}) & = &  {i \over {4 \pi \tau}} \int_M \left({\bf W}^+ - 2 \pi \eta \delta^+_D\right) \wedge \star \left({\bf W}^+ - 2 \pi \eta \delta^+_D\right) - {i \over {4 \pi \bar\tau}} \int_M \left({\bf W}^- - 2 \pi \eta \delta^-_D\right)\wedge \star \left({\bf W}^- - 2 \pi \eta \delta^-_D\right) \nonumber \\
&& - i \alpha \int_D ({\bf W} - 2 \pi \eta \delta_D).
\label{2.23}
\end{eqnarray}

Finally, by comparing the action ${\widehat {\bf I}} ({\bf w})$ with its $\it{equivalent}$ action
\begin{eqnarray}
({\bf{I}}'_{\tau} + {\bf{I}}_{\eta})(A)  & = & -{{i \tau} \over 4\pi} \int_M (F^{+} - 2\pi \alpha \delta^{+}_D) \wedge \star (F^{+}- 2\pi \alpha \delta^{+}_D) + {{i \bar\tau} \over 4\pi} \int_M (F^{-} - 2\pi \alpha \delta^{-}_D) \wedge \star (F^{-} - 2\pi \alpha \delta^{-}_D) \nonumber \\
&&  + i \eta \int_D (F - 2\pi \alpha \delta_D),
\end{eqnarray}
we see that the original theory in the gauge field $A$ with complexified coupling parameter $\tau$ and surface operator parameters $(\alpha, \eta)$, is $\it{dual}$ to a theory in the gauge field $\bf{w}$ with complexified coupling parameter $-1 / \tau$ and surface operator parameters $(\eta, - \alpha)$. In other words, we have explicitly shown that the pure $U(1)$ gauge theory with a trivially-embedded surface operator continues to enjoy $S$-duality, whereby the surface operator parameters transform as
\be
(\alpha, \eta) \to (\eta, - \alpha)
\label{invert}
\ee
under the $S$-duality transformation $S:\tau \to - 1/ \tau$.

\vspace{0.3cm}

\smallskip\noindent{\it Action Under a Shift in Theta-Angle}

Our second objective is to prove that the parameters $(\alpha, \eta)$ transform under the symmetry $T: \tau \to \tau + 1$ as
\begin{eqnarray}
\eta & \to & \eta - \alpha \nonumber \\
\alpha & \to & \alpha,
\end{eqnarray}
for $M$ a closed $\it{spin}$ manifold.

Note at this point that a proof of the above claim can also be found in $\S$2.5 of~\cite{Gukov-Witten}. However, there are some minor but non-trivial differences in our approaches, which nevertheless lead to the same conclusion.

Coming back to our discussion, note that the theta-angle term from our effective action $\bf{I}'_{\tau}$ in (\ref{I'tau}) is given by
\be
{\bf I}'_{\theta} = -{i \theta \over {8 \pi^2}} \int_M  F' \wedge F',
\ee
where $F' = F - 2 \pi \alpha \delta_D$. This can also be written as
\be
{\bf I}'_{\theta}  = - i \theta {\bf N},
\ee
where
\be
{\bf N} = {1 \over 2} c_1({\cal L})^2  - \alpha \frak m,
\label{N}
\ee
    and $\frak m = \int_D (F/ 2\pi)$ is the ``magnetic charge'' associated with the flux through $D$. A term $(\alpha^2 / 2) D \cap D$ that generically appears in $\bf N$ has been set to zero above, as we are considering surface operators which are trivially-embedded only. Also, the first term $(1/2) c_1({\cal L})^2$ is always an integer, since $M$ is defined to be spin.

Next, consider the term
\be
{\bf I}_{\eta}  = i \eta \int_D F' = i \eta \int_M \delta_D \wedge F',
\ee
which is the only term in the total action that can potentially cancel the variation of ${\bf I}'_{\theta}$ under the transformation $T: \theta \to \theta + 2 \pi$. It can also be written as
\be
{\bf I}_{\eta}  =  2 \pi i \eta {\frak m},
\label{I_shift}
\ee
where a term $- 2 \pi i \alpha \eta D \cap D$ has been set to zero in ${\bf I}_{\eta}$ above, since we are considering only trivially-embedded surface operators.

Thus, the sum of the two contributions to the total action is then
\be
{\bf I}'_{\theta} + {\bf I}_{\eta} = - i \theta {\bf N} + 2 \pi i \eta {\frak m}.
\ee

The variation in ${\bf I}'_{\theta}$ under $T: \theta \to \theta + 2 \pi$ is (mod $2 \pi i \mathbb Z$)
\be
\Delta {\bf I}'_{\theta} =   2 \pi i \alpha \frak m.
\ee
Hence, in order for the total contribution ${\bf I}'_{\theta} + {\bf I}_{\eta}$ to be invariant, one must have the transformation
\begin{eqnarray}
\eta & \to & \eta - \alpha\nonumber \\
\alpha &\to & \alpha
\label{shift}
\end{eqnarray}
under $T:\tau \to \tau +1$, as claimed.

If $M$ is $\it{not}$ a spin manifold, the original theory without surface operators is only invariant under $T^2: \tau \to \tau +2$. This is because $c_1({\cal L})^2$ is no longer an even integer. Repeating the above analysis, we find that the parameters must transform as
\begin{eqnarray}
\eta & \to & \eta - 2 \alpha\nonumber \\
\alpha &\to & \alpha
\label{shift non-spin}
\end{eqnarray}
under $T^2:\tau \to \tau +2$, when $M$ is not spin.

\vspace{0.3cm}

\smallskip\noindent{\it Action Under Overall Duality}

Note that the $SL(2,\mathbb Z)$ duality group is an infinite discrete group which acts on $\tau$ as
\be
\tau \to {{(a \tau + b)} \over {(c \tau +d)}},  \qquad \left(\begin{array}{ccc} a & & b \\ c & & d \end{array} \right) \in SL(2,\mathbb Z).
\ee
It is generated by the transformations $S: \tau \to -1 /\tau$ and $T: \tau \to \tau +1$, where
\be
S = \left(\begin{array}{ccc} 0 & & 1 \\ -1 & & 0 \end{array} \right), \qquad T = \left(\begin{array}{ccc} 1 & & 1 \\ 0 & & 1 \end{array} \right).
\ee
From (\ref{invert}) and (\ref{shift}), we find that $\alpha$ and $\eta$ transform as
\be
(\alpha, \eta) \to (\alpha, \eta) {\cal M}^{-1},
\ee
where $\cal M$ is $S$ or $T$, accordingly. Therefore, this is true for any ${\cal M} \in SL(2,\mathbb Z)$. Hence, we see that $(\alpha, \eta)$ transform naturally under the $SL(2,\mathbb Z)$ duality of the pure $U(1)$ gauge theory on a (closed) spin manifold $M$. In particular, $(\alpha, \eta)$ transform under $S$-duality just like magnetic and electric charge, respectively.

On the other hand, consider the congruence subgroup $\Gamma_0(2)$ that is generated by the transformations $S$ and $ST^2S$, that is,
\be
S = \left(\begin{array}{ccc} 0 & & 1 \\ -1 & & 0 \end{array} \right), \qquad ST^2S = \left(\begin{array}{ccc} -1 & & 0 \\ 2 & & -1 \end{array} \right).
\ee
From (\ref{invert}) and (\ref{shift non-spin}), we find that $\alpha$ and $\eta$ transform as
\be
(\alpha, \eta) \to (\alpha, \eta) {\cal M}^{'-1},
\ee
where ${\cal M}'$ is $S$ or $ST^2S$, accordingly. Therefore, this is true for any ${\cal M}' \in \Gamma_0(2)$. Hence, we see that $(\alpha, \eta)$ transform naturally under a $\Gamma_0(2)$ duality of the pure $U(1)$ gauge theory on a (closed) non-spin manifold $M$. Nonetheless, $(\alpha, \eta)$ continue to transform under $S$-duality just like magnetic and electric charge, respectively.

\newsubsection{Action of Duality on Non-Trivially-Embedded Surface Operators}

\smallskip\noindent{\it Action Under $S$-duality}

The analysis for a non-trivially-embedded surface operator is similar to the one before for a trivially-embedded surface operator, except for one minor difference. Recall that for a non-trivially-embedded surface operator, we have an additional term of $-2 \pi i \eta \alpha \int_M \delta_D \wedge \delta_D$ in the extended action $\widehat {\bf I}({\bf g}, {\bf w})$ of (\ref{2.20}). Consequently, one also ends up with this additional term in the final expression for the extended action, that is, the final extended action is in this case
\be
{\widehat {\bf I}} ({\bf w})' = {\widehat {\bf I}} ({\bf w}) -2 \pi i \eta \alpha \int_M \delta_D \wedge \delta_D,
\label{action with non-invariance term}
\ee
where ${\widehat {\bf I}} ({\bf w})$ is given in (\ref{2.23}).

From (\ref{intersection number}), we learn that $\alpha \int_M \delta_D \wedge \delta_D = \alpha D \cap D$ must be an integer. Therefore, for ${\widehat {\bf I}} ({\bf w})'$ above to be equivalent to ${\widehat {\bf I}} ({\bf w})$ mod $2 \pi i \mathbb Z$, the parameter $\eta$ must also be integer-valued. That is,
\be
{\widehat {\bf I}} ({\bf w})' = {\widehat {\bf I}} ({\bf w}) \quad \textrm{mod}\ 2\pi i \mathbb Z, \qquad \forall \ \eta \in \mathbb Z.
\ee

Thus, starting with the original theory with action ${\bf I}'_\tau + {\bf I}_\eta ({A})$, one can only arrive at a dual theory in terms of the gauge field $\bf w$, which has inverted complexified coupling parameter $-1/\tau$ and surface operator parameters $(\eta, - \alpha)$, if and only if $\eta$ is integer-valued. In other words, $\it exact$ $S$-duality will only be preserved if $\eta$ is integer-valued.

\vspace{0.3cm}

\smallskip\noindent{\it Action Under a Shift in Theta-Angle}

Recall also that for a non-trivially-embedded surface operator, one must add the terms $(\alpha^2 / 2) D \cap D$ and $- 2 \pi i \alpha \eta D \cap D$ to (\ref{N}) and (\ref{I_shift}), respectively. Hence, we have in this case
\be
{\bf I}'_{\theta} + {\bf I}_\eta = -i \theta \left( {\bf N} + (\alpha^2 / 2) D \cap D\right) + 2 \pi i \eta \left(\frak m - \alpha D \cap D \right),
\ee
where $\bf {N}$ is given in (\ref{N}).

For $M$ spin, the variation in ${\bf I}'_{\theta}$ under $T: \theta \to \theta + 2 \pi$ is now (mod $2 \pi i \mathbb Z$)
\be
\Delta {\bf I}'_{\theta} = 2 \pi i \alpha \frak m - \alpha \pi i \mathbb Z,
\ee
where we have made use of the fact that $\alpha D \cap D \in \mathbb Z$.

Suppose we have the transformation
\begin{eqnarray}
\eta & \to & \eta - \alpha\nonumber \\
\alpha &\to & \alpha
\label{shift 2}
\end{eqnarray}
under $T:\tau \to \tau + 1$. Then, the corresponding variation in ${\bf I}_{\eta}$ will be given by
\be
\Delta {\bf I}_{\eta} = -2 \pi i \alpha \frak m + 2  \alpha \pi i \mathbb Z.
\ee

In order for the theory to be invariant under $T: \tau \to \tau +1$ when the parameters of the surface operator transform as in (\ref{shift 2}), we must have $\Delta {\bf I}'_{\theta} + \Delta {\bf I}_{\eta} = \alpha \pi i \mathbb Z = 0 \ \textrm{mod} \ 2\pi i \mathbb Z$. In other words, $\alpha$ can only be even-integer-valued, for $M$ spin.

For $M$ non-spin, the variation in ${\bf I}'_{\theta}$ under $T^2: \theta \to \theta + 4 \pi$ is now (mod $2 \pi i \mathbb Z$)
\be
\Delta {\bf I}'_{\theta} = 4 \pi i \alpha \frak m -  2\alpha \pi i \mathbb Z.
\ee
Suppose we have the transformation
\begin{eqnarray}
\eta & \to & \eta - 2\alpha\nonumber \\
\alpha &\to & \alpha
\label{shift 2 non-spin}
\end{eqnarray}
under $T^2:\tau \to \tau + 2$. Then, the corresponding variation in ${\bf I}_{\eta}$ will be given by
\be
\Delta {\bf I}_{\eta} = -4 \pi i \alpha \frak m + 4  \alpha \pi i \mathbb Z.
\ee

In order for the theory to be invariant under $T^2: \tau \to \tau +2$ when the parameters of the surface operator transform as in (\ref{shift 2 non-spin}), we must have $\Delta {\bf I}'_{\theta} + \Delta {\bf I}_{\eta} = 2 \alpha \pi i \mathbb Z = 0 \ \textrm{mod} \ 2\pi i \mathbb Z$. In other words, $\alpha$ can only be integer-valued, for $M$ non-spin.

\vspace{0.3cm}

\smallskip\noindent{\it Action Under Overall Duality}

Let us now summarise the action of the transformations $S$, $T$ and $ST^2S$ on the parameters $(\alpha, \eta)$. For
\be
S = \left(\begin{array}{ccc} 0 & & 1 \\ -1 & & 0 \end{array} \right), \qquad T = \left(\begin{array}{ccc} 1 & & 1 \\ 0 & & 1 \end{array} \right) \qquad ST^2S = \left(\begin{array}{ccc} -1 & & 0 \\ 2 & & -1 \end{array}\right),
\ee
We find that $\alpha$ and $\eta$ transform as
\be
(\alpha, \eta) \to (\alpha, \eta) {\cal M}^{-1},
\label{transform2}
\ee
where $\cal M$ is $S$ or $T$ for $M$ spin, or is $S$ or $ST^2S$ for $M$ non-spin. However, in contrast to the previous case of a trivially-embedded surface operator, $\eta$ and $\alpha$ have to be restricted to integer and even-integer values, respectively, for $M$ spin, and only integer values for $M$ non-spin, as explained above.

Recall at this point from $\S$2.1, that $\eta$, by definition, must take values in $\mathbb R / \mathbb Z$. Hence, taking $\eta$ to be integer-valued is equivalent to setting $\eta$ to zero. In other words, exact $S$-duality will only be preserved in the free, $U(1)$ gauge theory for a class of $\textrm{\it{non-trivially-embedded}}$ surface operators which effectively have parameters $(\alpha, \eta) = (\alpha, 0)$. Alternatively, notice that since the term $-2 \pi i \eta \mathbb Z$ that results in the non-invariance is a $c$-number independent of the quantum fields, one could instead allow $\eta$ to be non-vanishing and arbitrarily-valued, and claim that $S$-duality holds up to a $c$-number.

Recall from footnote~1 that $\alpha$ (the $S$-dual of $\eta$) takes values in $\mathbb R / \mathbb Z$ too. This means that $\alpha$ also effectively vanishes if it is any integer. Hence, it will mean that the condition $\Delta {\bf I}'_{\theta} + \Delta {\bf I}_{\eta} = \alpha \pi i \mathbb Z = 0 \ \textrm{mod} \ 2\pi i \mathbb Z$, or the condition $\Delta {\bf I}'_{\theta} + \Delta {\bf I}_{\eta} = 2\alpha \pi i \mathbb Z = 0 \ \textrm{mod} \ 2\pi i \mathbb Z$ - which ensures invariance of the theory under $T:\tau \to \tau +1$ or $T^2:\tau \to \tau +2$, respectively - cannot really be satisfied for any non-zero value of $\alpha$. At any rate, the term $\alpha \pi i \mathbb Z $ or $2\alpha \pi i \mathbb Z $, which results in the non-invariance, is a $c$-number independent of the quantum fields. Therefore, one could instead allow $\alpha$ to be non-vanishing and arbitrarily-valued, and claim that the symmetry $T: \tau \to \tau +1$ holds up to a $c$-number.

 Hence, we can conclude that for any non-trivially-embedded surface operator, the parameters $(\alpha, \eta)$ will transform naturally under $SL(2,\mathbb Z)$ (or $\Gamma_0(2)$ when $M$ is non-spin) as shown in (\ref{transform2}). However,  exact $S$-duality will only hold if $\eta$ is effectively zero, while the overall $SL(2,\mathbb Z)$ (or $\Gamma_0(2)$) duality holds up to a $c$-number at most, regardless.

\vspace{0.3cm}

\smallskip\noindent{\it The Theory over Parameter Space}

Last but not least, take notice of the term $-2 \pi i \eta \alpha D \cap D$ which appears in the action ${\widehat{\bf I}}(\bf w)'$ of (\ref{action with non-invariance term}). Now, recall that any lattice shift of $\alpha$ (induced by a gauge transformation) must respect the condition $\alpha D \cap D \in \mathbb Z$, and that any lattice shift in $\eta$ is a shift by an integer. Altogether, this means that ${\widehat{\bf I}}(\bf w)'$ is invariant mod $2\pi i \mathbb Z$ under lattice shifts of $\alpha$ and $\eta$. Hence, at every inequivalent point in the $(\alpha,\eta)$-parameter space, we have a single-valued partition function. In other words, the partition function - even when $D\cap D \neq 0$ - is a complex-valued function of $\alpha$ and $\eta$.

Note that our above statement differs from the conclusion in $\S$2.5 of~\cite{Gukov-Witten}, which asserts that the partition function is a section of a non-trivial complex line bundle over the $(\alpha,\eta)$-parameter space. This discrepancy can be understood as follows. In~\cite{Gukov-Witten}, an arbitrary $c$-number term $-\pi i \alpha \eta D \cap D$ has been added by hand to the action, so that $T$ would persist as a symmetry of the theory under (\ref{shift 2}) (or (\ref{shift 2 non-spin})) for any value of $\alpha$, not just for even integer (or integer) values, as we have found earlier. Without this additional term, the partition function in~\cite{Gukov-Witten} would be a complex-valued function of $\alpha$ and $\eta$. But the anomaly in the symmetry $T:\tau \to \tau +1$ would now, in their discussion, be given by $-\pi i \alpha \mathbb Z$; that is, (ignoring possible lattice shifts in $\alpha$), the authors of~\cite{Gukov-Witten} would have to insist that $\alpha$ take on even integer values if $T:\tau \to \tau +1$ were to remain a symmetry under (\ref{shift 2}); this is simply our result above.

One can also interpret the above facts as follows. Firstly, adding the $c$-number term $-\pi i \alpha \eta D \cap D$ in~\cite{Gukov-Witten} actually makes the theory anomalous, because the partition function is now no longer a function over parameter space, but a section of a non-trivial line bundle over it. However, at the expense of inheriting this anomaly, one would have an apparent symmetry $T:\tau \to \tau +1$ (or $T^2:\tau \to \tau +2$) under (\ref{shift 2}) (or (\ref{shift 2 non-spin})), for any value of $\alpha$. Nevertheless, if one restricts to a subspace of the $(\alpha, \eta)$-parameter space where $\alpha$ takes on even integer (or integer) values, the line bundle over it can be made trivial, and the theory will be genuinely anomaly-free (if one again ignores possible lattice shifts in $\alpha$). This just re-expresses the claim in~\cite{Gukov-Witten} that it would be possible to omit the $c$-number term, and have the symmetry $T:\tau \to \tau +1$ hold up to a $c$-number (which we know can be canceled mod $2\pi i \mathbb Z$ if one chooses even integer values of $\alpha$).

\newsubsection{Correlation Functions with Wilson and 't Hooft Loop Operators}

We shall now consider a simple example of a correlation function between a Wilson loop operator and a surface operator. Let the surface operator ${\cal O}_D$ be supported on $D = S^2$ in $M = \mathbb R^4$. Let $D$ be linked by a circle $\cal C$ of unit radius in a plane normal to $D$. Then, the semi-classical approximation of a correlator of ${\cal O}_D$ with a Wilson loop operator ${W}_{\cal C} = \textrm{exp}(-\oint_{\cal C} A)$, amounts to evaluating the Wilson loop operator on the gauge field produced by the surface operator in (\ref{A}). As such, we have
\be
{{\langle {\cal O}_D \cdot {W}_{\cal C} \rangle } \over {\langle {\cal O}_D \rangle}} = \textrm{exp} (- 2 \pi \alpha).
\label{correlation 1}
\ee

Next, let us consider the correlation function between an 't Hooft loop operator and the above surface operator. An 't Hooft loop can be represented by a Dirac monopole, and in the limit that the loop is infinitely large, the field strength around the monopole (in the classical approximation) is exactly given by
\be
F = -{i\over 2} \ \textrm{sin} \theta \ d\theta \ d \phi,
\ee
where $\theta$ and $\phi$ are local coordinates on any unit two-sphere which surrounds the loop. Now, let the surface operator ${\cal O}_D$ wrap the loop, that is, let $D$ be the unit two-sphere surrounding the loop. Then, the surface operator would couple to the magnetic field produced by the 't Hooft operator through the parameter $\eta$ in (\ref{eta term}). As such, the correlator would be given by
\be
\textrm{exp} \left(  -i \eta \int_{S^2}( -{i \over 2}\ \textrm{sin} \theta \ d\theta \ d \phi) \right) = \textrm{exp}(- 2\pi \eta).
\label{correlation 2}
\ee

Under $S$-duality, the Wilson loop becomes an 't Hooft loop, that is, (\ref{correlation 1}) will become (\ref{correlation 2}). This however, can be effected by $\alpha \to \eta$ also. Consequently, we find that the parameter transformation $\alpha \to \eta$ under $S$-duality, is indeed consistent with a switch from Wilson to 't Hooft loop operators.

\newsection{Deriving the Transformation of Parameters via the Formalism of Duality Walls}

\vspace{-0.0cm}

\newsubsection{The Formalism of Duality Walls}

It was recently shown in~\cite{Gaitto} that for any element $g$ of the $SL(2,\mathbb Z)$ duality group of the pure $U(1)$ gauge theory on a four-manifold, one can define a codimension one defect that separates the four-manifold into two regions, such that the theories defined in each region are related by the duality transformation $g$ effected by a wall operator placed along the defect. As such, this wall operator is also know as a duality wall. Since any element $g$ can be generated from the transformations $S$ and $T$, it suffices for us to describe the corresponding wall operators associated to $S$ and $T$.

 Let us first describe the wall operator associated to the $S$-transformation $\tau \to -1/\tau$. Suppose the codimension one defect $W$ splits $M$ into two regions $M_-$ and $M_+$. We will choose the orientation of $W$ such that it agrees with the one induced from $M_-$ and disagrees with the one induced from $M_+$. Let $A$ and $\hat A$ be the gauge field and its $S$-dual living in $M_-$ and $M_+$, respectively. Then, the wall operator associated to the $S$-transformation can be defined by inserting into the path integral the factor~\cite{Gaitto}
\be
\textrm{exp} \left( - {i \over {2 \pi}} \int_W A \wedge d{\hat A}  \right).
\label{S-wall}
\ee
In other words, one must add to the action, the term ${i \over {2 \pi}} \int_W A \wedge d{\hat A}$. Thus, the effective action of the theory in region $M_-$ is given by
\be
\label{I_eff_M-}
{\bf I}_{M_-}  =  {1\over g^2} \int_M F \wedge \star F - {{i \theta}\over 8 \pi^2}\int_M F \wedge F + {i \over {2 \pi}} \int_W A \wedge d{\hat A},
\ee
while the effective action of the $S$-$\it{dual}$ theory in region $M_+$ is given by
\be
\label{I_eff_M+}
{\bf I}_{M_+}  =  {1\over {\hat g}^2} \int_M {\hat F} \wedge \star {\hat F} - {{i {\hat \theta}}\over 8 \pi^2}\int_M {\hat F} \wedge {\hat F} - {i \over {2 \pi}} \int_W A \wedge d{\hat A},
\ee
where $\hat F = d \hat A$ and $\hat \tau = {\hat \theta}/2\pi + 4\pi i /{\hat g}^2$ are the $S$-dual field strength and complexified gauge coupling, respectively. The minus sign in the last term of ${\bf I}_{M_+}$ arises because the wall $W$ and $M_+$ are defined to have opposite orientations.

One can see that (\ref{S-wall}) indeed corresponds to an S-duality wall operator as follows. By varying the actions ${\bf I}_{M_-}$ and ${\bf I}_{M_+}$, and requiring that the resulting boundary terms vanish as well, we find the following respective conditions on the fields:
\begin{eqnarray}
\label{hat F}
{\hat F}|_W & = & {4 \pi i \over{g^2}} \star F|_W - {\theta \over {2 \pi}} F|_W \nonumber \\
{F}|_W & = & - {4 \pi i \over{{\hat g}^2}} \star {\hat F}|_W + {{\hat \theta} \over {2 \pi}} {\hat F}|_W.
\end{eqnarray}
Noting that the expression for the stress-energy tensor is given by
\be
T_{\mu \nu} = {2 \over g^2} \left( F_{\mu \alpha} F^{\alpha}_{\nu} + {1\over 4} g_{\mu\nu} F_{\alpha \beta} F^{\alpha \beta} \right),
\ee
and substituting $F$ and $\hat F$ from (\ref{hat F}) into the stress-energy tensors $T$ and $\hat T$ of the theories in $M_-$ and $M_+$, one finds that $T = \hat T$ if and only if
\be
\hat \tau = - {1 / \tau}.
\label{S-duality}
\ee
In other words, we find that the theories in $M_-$ and $M_+$ are equivalent and $S$-dual to each other, as anticipated.

Similarly, the wall operator associated to the $T$-transformation $\tau \to \tau +1$, can be defined by inserting into the path integral the factor
\be
\textrm{exp} \left( - {i \over 4 \pi} \int_W A \wedge dA \right).
\label{T-operator}
\ee
Notice that the term ${i \over 4 \pi} \int_W A \wedge dA $ which one must now add to the action is manifestly topological and independent of the metric. Hence, it does not contribute to the stress-energy tensor $T_{\mu \nu} = {\delta S / \delta g_{\mu\nu}}$. Consequently, the stress-energy tensor does not get modified in the presence of the operator (\ref{T-operator}), and in particular, the stress-energy tensors of the theories in $M_-$ and $M_+$ agree across $W$. That is, (\ref{T-operator}) indeed represents a duality wall operator.

The duality wall operators of (\ref{S-wall}) and (\ref{T-operator}) have recently been utilised in~\cite{Anton} to derive - among other things - the transformation properties of non-local operators such as the Wilson-'t Hooft loop operator and Chern-Simons operator, under the $SL(2,\mathbb Z)$ duality of the pure U(1) gauge theory. However, the analysis for surface operators is lacking in~\cite{Anton}, and we shall now attempt to bridge this gap.

\newsubsection{Transformation of Parameters Under $T: \tau \to \tau +1$}

Let us now derive the transformation of the surface operator parameters $(\alpha, \eta)$ under $T: \tau \to \tau +1$, via the formalism of duality walls.

In what follows, we shall, for simplicity, assume that the surface operator is trivially-embedded, that is, we shall take $M = D \times \mathbb C$ to be the spin four-manifold, where the surface operator is supported along $D$, at the origin of $\mathbb C$. Also, as mentioned earlier, $z = r e^{i\theta}$ shall be the coordinate on $\mathbb C$, such that $D$ lies along $z=0$. Let us then define $W$ to be the three-dimensional boundary $\partial {\cal Z}^{\epsilon}_D = D \times \cal C$ of a tubular neighbourhood ${\cal Z}^{\epsilon}_D = D \times B^2_{\epsilon}$ of the surface operator supported along $D$, with ``thickness'' $\epsilon$, where $B^2_{\epsilon}$ is a disc of unit radius $\epsilon$ centred at the origin of $\mathbb C$, with boundary $\partial B^2_{\epsilon}$ the circle $\cal C$.

By inserting the operator (\ref{T-operator}) into the path integral, we are effectively placing a duality wall along $W$, which will divide $M$ into the regions $M_-$ and $M_+$, that lie exteriorly and interiorly of ${\cal Z}^{\epsilon}_D$, respectively. Because the region $M_+$ (with opposite orientation from $W$) contains the surface operator supported along $D$, the additional term $-{i \over 4 \pi} \int_{\partial {\cal Z}^{\epsilon}_D} A \wedge dA$ which now appears in the action of the theory in $M_+$, must be evaluated on the gauge field produced by the surface operator itself. In particular, recall that for the circle ${\cal C}$ linking $D$, we have $\oint_{\cal C} A = 2 \pi \alpha$, since $A = \alpha d \theta$ in $\mathbb C$. Thus, the additional term that appears in the action of the $\it{dual}$ theory in $M_+$, will be given by\footnote{We have taken advantage of the fact that the additional term $-{i \over 4 \pi} \int_{\partial {\cal Z}^{\epsilon}_D} A \wedge dA$ is manifestly topological and independent of the metric on $\partial {\cal Z}^{\epsilon}_D$, and rescaled the radius $\epsilon$ appropriately.}
\be
-i \alpha \int_{D} F.
\label{together}
\ee
Since the definition (\ref{eta term}) of a surface operator requires one to include in the original action the term $i \eta \int_D F$, and since the insertion of the operator (\ref{T-operator}) in the path integral does not modify the gauge field $A$, we find that together with (\ref{together}), we will have $(\alpha, \eta) \to (\alpha, \eta -\alpha)$ under $T : \tau \to \tau +1$, as proven earlier.

\newsubsection{Transformation of Parameters Under $S: \tau \to - 1 /\tau$}

We shall now derive the transformation of the surface operator parameters $(\alpha, \eta)$ under $S: \tau \to - 1/ \tau$, via the formalism of duality walls. Again, we shall assume that the surface operator is trivially-embedded.

As a start, recall that the definition of a surface operator requires one to insert into the path integral the factor
\be
\textrm{exp} \left(- i \eta \int_D F \right).
\ee
Notice that we can also write this as
\be
\textrm{exp} \left ({i\over 2\pi} \int_{\partial {\cal Z}^{\epsilon}_D} F \wedge \Omega_{\eta} \right)
\label{surface operator insert}
\ee
for a one-form $\Omega_{\eta}$ on $M$ that obeys $\oint_{\cal C} \Omega_{\eta} = - 2\pi \eta$, where $\eta \in \mathbb R /\mathbb Z$.

Does the one-form $\Omega_{\eta}$ exist, one may ask. To answer this question, first notice that our above definition asserting that $D$ includes only the centre $z =0$ of $B^2_{\epsilon}$, implies that ${\cal C}$ (that is, $\partial B^2_{\epsilon}$) and $D$ have linking number one. Consequently, this means that $D$ must be homologically trivial, that is, $D$ must be a boundary of a three-chain $D^3$, such that the algebraic intersection number of $D^3$ and $\cal C$ is one~\cite{Bott}. That $D$ must be homologically trivial is in fact consistent with a required property of the term $i \eta \int_D F$ which defines the surface operator: $\int_D F/ 2\pi$ is the integrated first Chern class of the $U(1)$-bundle over $D$. As such, the term $i \eta \int_D F$ ought to be invariant under the shift $F \to F + df$, where $f$ is some globally-defined one-form on $D$. This is only possible if $D$ is homologically trivial with no boundary. That the intersection number of $D^3$ and $\cal C$ is one, simply means that we have $\int_M \Omega_{D^3} \wedge \Omega_{\cal C} = \oint_{\cal C} \Omega_{D^3} = 1$, where the one-form $\Omega_{D^3}$ and the three-form $\Omega_{\cal C}$ are the Poincar\'e-dual classes of $D^3$ and $\cal C$ in $M$, respectively. Hence, the one-form $\Omega_{\eta}$ does indeed exist, and is given by  $\Omega_{\eta}= -2\pi \eta \cdot \Omega_{D^3}$.

Let us now place the duality wall (\ref{S-wall}) along $W = \partial {\cal Z}^{\epsilon}_D$. Together with the factor (\ref{surface operator insert}), and noting that $\partial W = 0$, we find that this is equivalent to inserting in the path integral the wall operator
\be
\textrm{exp} \left (-{i\over 2\pi} \int_{\partial {\cal Z}^{\epsilon}_D} A \wedge d\hat B \right),
\label{wall operator}
\ee
where $\hat B = \hat A - \Omega_{\eta}$.\footnote{We have - in deriving (\ref{wall operator}) - made use of the fact that the requisite conditions in (\ref{hat F}) admit a solution whereby $F|_{W = \partial {\cal Z}^{\epsilon}_D}$ is trivial in cohomology, such that one can write $F=dA$ globally over $\partial {\cal Z}^{\epsilon}_D$.} 
Notice that this is just an $S$-duality wall operator. This means that the $S$-dual theory in $M_+$ within $\partial {\cal Z}^{\epsilon}_D$ containing the surface operator, has dual gauge field $\hat B$ and complexified gauge coupling $\hat \tau = - 1 /\tau$. Its action will also contain the extra boundary term $-{i\over 2\pi} \int_{\partial {\cal Z}^{\epsilon}_D} A \wedge d\hat B$ from the insertion (\ref{wall operator}).

Note at this point that the condition $\oint_{\cal C} \Omega_{\eta} = - 2\pi \eta$ implies that $\Omega_{\eta} = \eta  d \theta$ (where we have made use of the fact that ${\cal C} \subset \partial {\cal Z}^{\epsilon}_D$, and the fact that the orientation of the region $M_+$ - where the field $\hat B$ and therefore the one-form $\Omega_{\eta}$ is defined in - is opposite to that of $\partial {\cal Z}^{\epsilon}_D$). Evaluating (\ref{wall operator}) on the gauge field $A = \alpha d\theta$ produced by the surface operator, and noting that $d \Omega_{\eta} = 2 \pi \eta\delta_D$, we finally find the effective action of the $S$-dual theory in $M_+$ to be given by
\begin{eqnarray}
\label{IM+}
{\bf I}_{M_+} (\hat B) & = & {1\over {\hat g}^2} \int_M ({\hat F - 2 \pi \eta \delta_D}) \wedge \star ({\hat F - 2 \pi \eta \delta_D}) - {{i {\hat \theta}}\over 8 \pi^2}\int_M ({\hat F - 2 \pi \eta \delta_D}) \wedge ({\hat F - 2 \pi \eta \delta_D}) \nonumber \\
&& - i \alpha \int_D (\hat F - 2 \pi \eta \delta_D).
\end{eqnarray}
This can be re-written as ${\widehat {\bf I}} ({\bf w})$ of (\ref{2.23}), if one replaces $\bf w$ with $\hat A$, and $\tau$ with $- 1 / \hat \tau$. In other words, under the $S$-duality transformation $S: \tau \to - 1 / \tau$, the surface operator parameters transform as $(\alpha, \eta) \to (\eta, - \alpha)$, as proven earlier.

As a final comment, notice that the one-form $\Omega_{D^3}$ and therefore the one-form $\Omega_{\eta}$, appears to be non-unique. This is because the Poincar\'e-dual three-chain $D^3$ is defined modulo an addition of a three-cycle. However, this extra degree of freedom will be fixed once $\Omega_{\eta}$ is required to satisfy a certain condition. Indeed, note that the equation of motion for the theory at the boundary $\partial {\cal Z}^{\epsilon}_D$ is, via (\ref{wall operator}), given by $d\hat B = 0$. This implies that $\Omega_{\eta}$ must satisfy the classical condition $\int_D \hat F = \int_D d \Omega_{\eta}$, and is therefore unique. In turn, since $d \Omega_{\eta} = 2 \pi \eta \delta_D$, and since $\delta_D$ is a delta two-form with support along $D$ only, it must be true that $\hat F = 2\pi \eta \delta_D + \dots$, where the ellipses refer to terms that are regular near $D$. In other words, $\hat F$ has the same form as $\bf W$ in (\ref{2.23}), such that ${\bf I}_{M_+} (\hat B)$ - like ${\widehat {\bf I}} ({\bf w})$ - is a non-divergent action with a non-zero contribution to the path integral, as required physically.

\newsection{Partition Function of Pure $U(1)$ Gauge Theory with Surface Operators}

\vspace{-0.0cm}

\newsubsection{Modular Forms}

For a function $F$ that is not necessarily holomorphic, we say that it transforms as a modular form of weight $(u,v)$ for a finite index subgroup $\Gamma$ of $SL(2,\mathbb Z)$ if for
\be
\left(\begin{array}{ccc} a & & b \\ c & & d \end{array} \right) \in \Gamma,
\ee
one has
\be
F\left( {{a\tau + b} \over {c\tau +d}} \right) = (c \tau + d)^u (c \bar \tau + d)^v F(\tau).
\ee

In particular, if we have a function $F(\tau)$ which transforms as
\begin{eqnarray}
\label{modular S-tx}
F(- 1 /\tau) & = & \tau^u {\bar \tau}^v Z(\tau) \\
F(\tau +1 ) & = & Z(\tau),
\end{eqnarray}
we say that $F(\tau)$ transforms as a modular form of $SL(2,\mathbb Z)$ with weight $(u,v)$.\footnote{This statement is to hold up to a $\tau$-independent multiplicative constant.} Alternatively, if $F(\tau)$ transforms as
\begin{eqnarray}
\label{modular S-tx non-spin}
F(- 1 /\tau) & = & \tau^u {\bar \tau}^v Z(\tau) \\
F(\tau +2) & = & Z(\tau),
\end{eqnarray}
we say that $F(\tau)$ transforms as a modular form of $\Gamma_0(2)$ with weight $(u,v)$.

We would like to extend the above definitions to functions which depend also on the parameters $\alpha$ and $\eta$. In light of the way $(\alpha, \eta)$ transform naturally under $SL(2,\mathbb Z)$ (or $\Gamma_0(2)$), we shall say that if $F(\tau, \alpha, \eta)$ transforms as
\begin{eqnarray}
\label{modular S-tx extended}
F(- 1 /\tau, \eta, -\alpha) & = & \tau^u {\bar \tau}^v Z(\tau, \alpha, \eta) \\
F(\tau +1, \alpha, \eta - \alpha) & = & Z(\tau, \alpha, \eta),
\end{eqnarray}
then it transforms like a modular form of $SL(2,\mathbb Z)$ with weight $(u,v)$. Alternatively, if $F(\tau, \alpha, \eta)$ transforms as
\begin{eqnarray}
\label{modular S-tx non-spin}
F(- 1 /\tau,\eta, -\alpha) & = & \tau^u {\bar \tau}^v Z(\tau,\alpha, \eta) \\
F(\tau + 2,\alpha, \eta - 2\alpha) & = & Z(\tau, \alpha, \eta),
\end{eqnarray}
we say that $F(\tau)$ transforms like a modular form of $\Gamma_0(2)$ with weight $(u,v)$.

\newsubsection{Partition Function of Pure $U(1)$ Gauge Theory with Surface Operators}

\smallskip\noindent{\it Partition Function with Trivially-Embedded Surface Operators}

The partition function of the original theory in the gauge field $A$ with complexified coupling parameter $\tau$, can be written as
\be
Z(\tau, \alpha, \eta) = (\textrm{Im}\tau)^{1/2(B_1 - B_0)} {1\over {\textrm{vol}({\cal G})}} \sum_{\cal L} \int {\cal D} A \ e^{- ({\bf I}'_{\tau} + {\bf I}_{\eta})(A, \tau, \alpha, \eta)},
\label{Z_tau}
\ee
where ${\bf I}'_{\tau} (A, \tau, \alpha, \eta)$ and ${\bf I}_{\eta}(A, \tau, \alpha, \eta)$ are given by (\ref{I'tau}) and (\ref{I_eta}), respectively, and $B_k$ denotes the dimension of the space of $k$-forms on $M$. The prefactor of $(\textrm{Im}\tau)^{1/2(B_1 - B_0)}$ arises because we have implicitly assumed a lattice regularisation of the path integral. In a lattice regularisation, one would include in the definition of the path integral a factor of $\textrm{Im} \tau^{1/2}$ for every integration variable, and a factor of $\textrm{Im} \tau^{-1/2}$ for every generator of a gauge transformation, so as to cancel a cut-off dependent factor. Since the integration variable $A$ and the gauge parameter $\epsilon$ (which generates the gauge transformation) is a one-form and zero-form on $M$, their numbers will be given by $B_1$ and $B_0$.  Thus, we have the resulting prefactor. $B_1$ and $B_0$ are of course infinite, but they can be made finite via the regularisation. We shall elaborate further on this point in a while.

Since as demonstrated earlier in $\S$2.2, the theory with action $\widehat{\bf I}(A, {\bf g}, {\bf w}, \tau, \alpha, \eta)$ in (\ref{extended action}) is equivalent to the original theory, we can also write the partition function as
\be
Z(\tau, \alpha, \eta) = (\textrm{Im}\tau)^{1/2(B_1 - B_0)} {1\over {\textrm{vol}({\cal G})}} {1\over {\textrm{vol}({\widehat {\cal G}})}} {1\over {\textrm{vol}({\widetilde {\cal G}})}} \sum_{{\cal L},{\widetilde{\cal L}}} \int {\cal D} A {\cal D} {\bf g} {\cal D} {\bf w} \ e^{- \widehat{\bf I}(A, {\bf g}, {\bf w}, \tau, \alpha, \eta)},
\label{Z}
\ee
where one recalls that $\widehat{\bf I}(A, {\bf g}, {\bf w}, \alpha, \eta)$ is explicitly given by
\be
{\widehat {\bf I}} (A, {\bf g}, {\bf w}, \alpha, \eta) = {i \over 2 \pi} \int_M {\bf W} \wedge {\bf g} - {{i \tau} \over 4\pi} \int_M {\cal F'}^{+} \wedge \star {\cal F'}^{+} + {{i \bar\tau} \over 4\pi} \int_M {\cal F'}^{-} \wedge \star {\cal F'}^{-} +i \eta \int_D {\cal F}'.
\label{extended action repeat}
\ee
Notice that the $\bf w$-dependent part of ${\widehat {\bf I}} (A, {\bf g}, {\bf w}, \alpha, \eta)$ is independent of $\tau$, and upon evaluating the $\bf w$-integral, one gets a $\tau$-independent delta function as explained earlier. One can then evaluate the $\bf g$-integral via this delta function without generating any powers of $\textrm{Im} \tau$. What is left behind then is just the integration variable $A$. As such, the prefactor in (\ref{Z}) is the same as that in (\ref{Z_tau}).

Alternatively, let us now evaluate (\ref{extended action repeat}) by gauging $A$ to zero. According to our computations in $\S$2.2, we can now write the partition function as
\be
Z(\tau, \alpha, \eta) = (\textrm{Im}\tau)^{1/2(B_1 - B_0)} {1\over {\textrm{vol}({\widehat {\cal G}})}} {1\over {\textrm{vol}({\widetilde {\cal G}})}} \sum_{\widetilde{\cal L}} \int {\cal D} {\bf g}' {\cal D} {\bf w} \ e^{- \widehat{\bf I}({\bf g}', {\bf w}, \tau, \alpha, \eta)},
\label{Z in g'}
\ee
where one recalls that $\widehat{\bf I}({\bf g}', {\bf w}, \tau, \alpha, \eta)$ is explicitly given by
\begin{eqnarray}
\label{I g' repeat}
{\widehat {\bf I}} ({\bf g}', {\bf w}, \tau, \alpha, \eta) & = & -{{i \tau} \over {4 \pi}} \int_M |{\bf g}^{'+}|^2 + {{i \bar\tau} \over {4 \pi}} \int_M |{\bf g}^{'-}|^2  + {i \over {4 \pi \tau}} \int_M |{\bf W}^+ - 2 \pi \eta \delta^+_D|^2 \nonumber \\
&& - {i \over {4 \pi \bar\tau}} \int_M |{\bf W}^- - 2 \pi \eta \delta^-_D|^2 - i \alpha \int_D ({\bf W} - 2 \pi \eta \delta_D).
\end{eqnarray}
In an eigenfunction expansion of ${\bf g}^{'+}$ and ${\bf g}^{'-}$, there are $B^+_2$ and $B^-_2$ modes  for ${\bf g}^{'+}$ and ${\bf g}^{'-}$, respectively, where $B^{\pm}_2$ are the dimensions of self-dual and anti-self-dual two-forms on $M$. From the $\tau$-dependence of the ${\bf g}'$-dependent terms in (\ref{I g' repeat}), it is clear that in evaluating the Gaussian integral over ${\bf g}'$ in (\ref{Z in g'}), one gets a factor of $({-i\tau/4 \pi})^{-1/2}$ and $({i\bar\tau / 4 \pi})^{-1/2}$ for every mode of ${\bf g}^{'+}$ and ${\bf g}^{'-}$. That is, we get a factor of
\be
\left({-i\tau \over 4 \pi}\right)^{-B^+_2 / 2}\left({i\bar\tau \over 4\pi}\right)^{-B^-_2 / 2}
\ee
after integrating over ${\bf g}'$ in (\ref{Z in g'}). Consequently, since ${\widehat {\bf I}} (0, {\bf w}, \tau, \alpha, \eta) = ({\bf I}'_{\tau} + {\bf I}_{\eta})({\bf w}, -1/\tau, \eta, -\alpha)$, we can write
\be
Z(\tau, \alpha, \eta) = (\textrm{Im}\tau)^{1/2(B_1 - B_0)} \tau^{-B^+_2 / 2} {\bar\tau}^{-B^-_2 / 2} {1\over {\textrm{vol}({\widetilde {\cal G}})}} \sum_{\widetilde{\cal L}} \int {\cal D} {\bf w} \ e^{- ({\bf I}'_{\tau} + {\bf I}_{\eta})({\bf w}, -1/\tau, \eta, -\alpha)}
\label{Z-dual}
\ee
up to a $\tau$-independent multiplicative constant. By noting that $\textrm{Im} (- 1 /\tau) = \textrm{Im}(\tau) / (\tau \bar \tau)$, and then by comparing with (\ref{Z_tau}) written in terms of the dual variable $\bf w$, we find that
\be
Z(\tau, \alpha, \eta) = \tau^{-{1\over 2}(B^+_2 - B_1 + B_0)}  {\bar\tau}^{-{1\over 2}(B^-_2 - B_1 + B_0)}  Z(-1/\tau, \eta, -\alpha). \label{Z-Z}
\ee
$B_2$, $B_1$ and $B_0$ are infinite, but can be made finite after the partition function is appropriately regularised. In the limit that $(\alpha, \eta) \to (0,0)$, we have no surface operators,\footnote{This statement however, is not true for surface operators in the twisted ${\cal N}=4$ SYM theory of~\cite{Gukov-Witten}; in the twisted ${\cal N}=4$ SYM theory, the expression of $A$ as given by (\ref{A}), is only defined modulo terms which are independent of $\alpha$ and less singular than $1/r$.} that is, we are back to the ordinary Maxwell theory studied in~\cite{abelian S-duality}. In order for our result in (\ref{Z-Z}) to agree with that in~\cite{abelian S-duality} when $(\alpha, \eta) \to (0,0)$, one must set $B_2 = b_2$, $B_1 = b_1$ and $B_0 = b_0$,\footnote{Note that due to a sign difference in our definition of the theta-term in the Lagrangian, one must switch $\tau \leftrightarrow -\bar \tau$ when comparing our results with that of~\cite{abelian S-duality}.} where $b_i$ is the $i$-th Betti number of $M$. Thus, since $b_0 - b_1 + b_2^{\pm} = (\chi \pm \sigma) / 2$, where $\chi$ and $\sigma$ are the Euler number and signature of $M$, respectively, we finally have
 \be
Z(-1/\tau, \eta, -\alpha) = \tau^{{1\over 4}(\chi + \sigma)}  {\bar\tau}^{{1\over 4}(\chi - \sigma)}Z(\tau, \alpha, \eta). \label{Z final}
\ee
Together with the fact that $({\bf I}'_{\tau} + {\bf I}_{\eta})(A, \tau +1, \alpha, \eta - \alpha) = ({\bf I}'_{\tau} + {\bf I}_{\eta})(A, \tau, \alpha, \eta)$ for $M$ spin, that is,
\be
Z(\tau + 1, \alpha, \eta - \alpha) = Z(\tau, \alpha, \eta),
\ee
we conclude that for any trivially-embedded surface operator, $Z(\tau, \alpha, \eta)$ transforms like a modular form of $SL(2, \mathbb Z)$ of weight $((\chi + \sigma) / 4, (\chi - \sigma)/ 4)$, when $M$ is spin.

For $M$ non-spin, we have instead $({\bf I}'_{\tau} + {\bf I}_{\eta})(A, \tau +2, \alpha, \eta - 2\alpha) = ({\bf I}'_{\tau} + {\bf I}_{\eta})(A, \tau, \alpha, \eta)$, that is,
\be
Z(\tau + 2, \alpha, \eta - 2\alpha) = Z(\tau, \alpha, \eta).
\ee
Hence, we conclude that for any trivially-embedded surface operator, $Z(\tau, \alpha, \eta)$ transforms like a modular form of $\Gamma_0(2)$ of weight $((\chi + \sigma) / 4, (\chi - \sigma)/ 4)$, when $M$ is non-spin.

\vspace{0.3cm}

\smallskip\noindent{\it Partition Function with Non-Trivially-Embedded Surface Operators}

The computation of the partition function for when the surface operator is non-trivially-embedded, is almost identical to the one above for when the surface operator is trivially-embedded, except for a minor difference.

When a surface operator is non-trivially-embedded, according to our discussion in $\S$2.3, one must add to $\widehat{\bf I}({\bf g}, {\bf w}, \tau, \alpha, \eta)$ that appears in the exponent in (\ref{Z in g'}), the $c$-number term $- 2 \pi i \eta \mathbb Z$. Being a $c$-number term that is independent of the quantum fields, it can be factored out of the path integral. One then proceeds as before to evaluate the Gaussian integral over the ${\bf g}'$ fields, which again yields a factor of $({-i\tau/4 \pi})^{-B_2^+/2}$ and $({i\bar\tau / 4 \pi})^{-B^-_2/2}$. Since one must replace $B_0$, $B_1$ and $B_2$ in (\ref{Z in g'}) by $b_0$, $b_1$ and $b_2$ under the appropriate regularisation, and since the relation ${\widehat {\bf I}} (0, {\bf w}, \tau, \alpha, \eta) = ({\bf I}'_{\tau} + {\bf I}_{\eta})({\bf w}, -1/\tau, \eta, -\alpha)$ remains unchanged, we will again have, after replacing $\bf w$ with the dual variable $A$, the result
\be
 Z(-1/\tau, \eta, -\alpha) = \tau^{{1\over 4}(\chi + \sigma)}  {\bar\tau}^{{1\over 4}(\chi - \sigma)}Z(\tau, \alpha, \eta), \label{Z final repeat}
\ee
up to a $\tau$-independent $c$-number.

Next, note also from our discussion in $\S$2.3 that for $M$ spin, the original action ${\bf I}'_{\tau} + {\bf I}_{\eta} (A)$ is invariant mod $2\pi i \mathbb Z$ under $T: \tau \to \tau +1$, up to a $c$-number term $\alpha \pi i \mathbb Z$. In other words, we have, for $M$ spin,
\be
Z(\tau + 1, \alpha, \eta - \alpha) = Z(\tau, \alpha, \eta)
\ee
up to a $\tau$-independent $c$-number.

In the case when $M$ is non-spin, the original action ${\bf I}'_{\tau} + {\bf I}_{\eta} (A)$ is invariant mod $2\pi i \mathbb Z$ under $T: \tau \to \tau +2$, up to a $c$-number term $2 \alpha \pi i \mathbb Z$. In other words, we have, for $M$ non-spin,
\be
Z(\tau + 2, \alpha, \eta - \alpha) = Z(\tau, \alpha, \eta)
\ee
up to a $\tau$-independent $c$-number also.

Recall from our discussion at the end of $\S$2.3 regarding the theory over parameter space, that the partition function continues to be a complex-valued function of $\alpha$ and $\eta$ even when the surface operators are non-trivially-embedded. Hence, altogether, we can conclude that for $M$ spin (or non-spin), the partition function of the pure $U(1)$ gauge theory with a non-trivially embedded surface operator - like the one with a trivially-embedded surface operator - transforms like a modular form of $SL(2,\mathbb Z)$ (or $\Gamma_0(2)$) of weight $((\chi + \sigma) / 4, (\chi - \sigma)/ 4)$. As required, our results reduce to that of the ordinary Maxwell theory in the limit $(\alpha, \eta) \to (0,0)$, where there are no surface operators.

\vspace{0.3cm}

\smallskip\noindent{\it Discussion on the Modular Anomaly}

Whenever $M$ is non-flat, $\chi$ and $\sigma$ do not vanish, and the partition function always transforms as a modular form rather than a modular-invariant function. One can interpret this observation as follows. When the pure $U(1)$ gauge theory is coupled to non-dynamical gravity (that is, to a fixed curved space), $S$-duality can only be maintained if one makes certain minimal $c$-number couplings that involve the background gravitational field. In light of (\ref{Z final}) (or (\ref{Z final repeat})), we find that the couplings involved must take the form
\be
- \int_M \left( \{B(\tau) + B(\bar \tau)\} \ \textrm{tr} R \wedge \widetilde R + \{C(\tau) + C(\bar \tau)\} \ \textrm{tr} R \wedge R\right),
\label{coupling}
\ee
where $\textrm{tr} R \wedge \widetilde R$ and $\textrm{tr} R \wedge R$ are the densities whose integrals give the Euler characteristic $\chi$ and signature $\sigma$ of $M$, respectively, and
\be
B(\tau)  =  C(\tau)  =  {1\over 8} \textrm{ln}(\tau^3 - {1\over \tau}), \qquad B(\bar \tau)  =  C(-{1 \over {\bar \tau}})  =  {1\over 8} \textrm{ln}({\bar\tau}^3 - {1\over {\bar \tau}}).
\ee

Note however, that even though one can - by adding the coupling term of (\ref{coupling}) to the original action - maintain $S$-duality on a curved four-manifold $M$, that is, we now have $Z(-1/\tau, \eta, -\alpha) =  Z(\tau, \alpha, \eta)$, one can no longer maintain - not even up to a $\tau$-independent $c$-number - the relation $Z(\tau + 1, \alpha, \eta - \alpha) = Z(\tau, \alpha, \eta)$ or $Z(\tau + 2, \alpha, \eta - \alpha) = Z(\tau, \alpha, \eta)$, when $M$ is spin or non-spin, respectively. In other words, the theory can never be fully modular-invariant in a curved four-manifold background.

\newsubsection{Transformation of Correlation Functions of Non-Singular, Gauge-Invariant Local Operators}

For positive integers $a$ and $b$, let us now consider a correlation function of an arbitrary monomial ${\cal O} (F'_+, F'_-) = (F'_+)^a(F'_-)^b$ of the $\it{non}$-$\it{singular}$, gauge-invariant local operators $F'_\pm = F_\pm - 2 \pi \alpha \delta^\pm_D$, in the background presence of an arbitrarily-embedded surface operator with parameters $(\alpha, \eta)$, at complexified gauge coupling $\tau$:
\be
{\langle  {\cal O} (F'_+, F'_-)  \rangle}_{\tau, \alpha, \eta} \sim (\textrm{Im}\tau)^{1/2(b_1 - b_0)} {1\over {\textrm{vol}({\cal G})}} \sum_{\cal L} \int {\cal D} A \ {\cal O} (F'_+, F'_-) \cdot  e^{- ({\bf I}'_{\tau} + {\bf I}_{\eta})(A, \tau, \alpha, \eta)}.
\label{corr of local operators 1}
\ee
We shall now evaluate how this correlation function transforms under $S$-duality, just like what was done above for the partition function.

To this end, first note that the correlation function (\ref{corr of local operators 1}) can also be computed as the following correlation function of the (equivalent) extended theory
\be
 (\textrm{Im}\tau)^{1/2(b_1 - b_0)} {1\over {\textrm{vol}({\cal G})}} {1\over {\textrm{vol}({\widehat {\cal G}})}} {1\over {\textrm{vol}({\widetilde {\cal G}})}} \sum_{{\cal L},{\widetilde{\cal L}}} \int {\cal D} A {\cal D} {\bf g} {\cal D} {\bf w} \ {\cal O} ({\cal F}'_+, {\cal F}'_-) \cdot e^{- \widehat{\bf I}(A, {\bf g}, {\bf w}, \tau, \alpha, \eta)},
\label{corr of local operators 2}
\ee
where one recalls that ${\cal F}'_{\pm} = F'_{\pm}- {\bf g}_{\pm}$.

Next, note that from (\ref{g'}), we can write ${\cal F}'_+ = F_+ - {\bf g}'_+ - ({1 / \tau}) ({\bf W}^+ - 2 \pi \eta \delta_D^+)$ and ${\cal F}'_- = F_ -  - {\bf g}'_- - ({1 / \bar\tau}) ({\bf W}^- - 2 \pi \eta \delta_D^-)$. By evaluating (\ref{corr of local operators 2}) in the gauge $A = 0$, and rewriting the action in terms of the field ${\bf g}'$ (see (\ref{I g'})), we have
\be
(\textrm{Im}\tau)^{1/2(b_1 - b_0)} {1\over {\textrm{vol}({\widehat {\cal G}})}} {1\over {\textrm{vol}({\widetilde {\cal G}})}} \sum_{\widetilde{\cal L}} \int {\cal D} {\bf g}' {\cal D} {\bf w} \ {\cal O} (\widehat {\cal F}'_+, \widehat {\cal F}'_-) \cdot e^{- \widehat{\bf I}({\bf g}', {\bf w}, \tau, \alpha, \eta)}
\label{corr of local operators 3}
\ee
up to a $\tau$-independent $c$-number, where $\widehat {\cal F}'_+ = - {\bf g}'_+ - ({1 / \tau}) ({\bf W}^+ - 2 \pi \eta \delta_D^+)$ and $\widehat {\cal F}'_- = - {\bf g}'_- - ({1 / \bar\tau}) ({\bf W}^- - 2 \pi \eta \delta_D^-)$.

Finally, by integrating out the fields ${\bf g}'_\pm$ from (\ref{corr of local operators 3}), and by noting that $\textrm{Im} (- 1 /\tau) = \textrm{Im}(\tau) / (\tau \bar \tau)$ and ${\widehat {\bf I}} (0, {\bf w}, \tau, \alpha, \eta) = ({\bf I}'_{\tau} + {\bf I}_{\eta})({\bf w}, -1/\tau, \eta, -\alpha)$, we get, up to a $\tau$-independent multiplicative constant
 \be
 \left(\textrm{Im} (- 1/\tau)\right)^{1/2(b_1 - b_0)} \tau^{- (a + {{\chi + \sigma} \over 4})}  {\bar\tau}^{-(b + {{\chi - \sigma} \over 4})} {1\over {\textrm{vol}({\widetilde {\cal G}})}} \sum_{\widetilde{\cal L}} \int {\cal D} {\bf w} \  {\cal O} ({\bf W}'_+, {\bf W}'_- ) \cdot e^{- ({\bf I}'_{\tau} + {\bf I}_{\eta})({\bf w}, -1/\tau, \eta, -\alpha)},
\label{corr of local operators 4}
\ee
where ${\bf W}'_\pm ={\bf W}^\pm - 2 \pi \eta \delta_D^\pm$.\footnote{We have, in the above computation, made use of the fact that the fields ${\bf g}'_\pm$ - as they appear in the explicit expression of $\widehat{\bf I}({\bf g}', {\bf w}, \tau, \alpha, \eta)$ in (\ref{I g' repeat}) - are non-propagating and can therefore be set to zero in ${\cal O} (\widehat {\cal F}'_+, \widehat {\cal F}'_-)$ via their classical equations of motion.} By comparing (\ref{corr of local operators 4}) with (\ref{corr of local operators 1}), we see that the correlation function ${\langle  {\cal O} (F'_+, F'_-)  \rangle}_{\tau, \alpha, \eta}$ transforms under $S$-duality as
\be
{\langle  {\cal O} (F'_+, F'_-)  \rangle}_{-1/\tau, \eta, -\alpha} = \tau^{(a + {{\chi + \sigma} \over 4})}  {\bar\tau}^{(b + {{\chi - \sigma} \over 4})}{\langle  {\cal O} (F'_+, F'_-)  \rangle}_{\tau, \alpha, \eta}
\ee
up to a $\tau$-independent $c$-number.

Since ${\cal O}(F'_+, F'_-)$ is independent of $\tau$, when $M$ is spin, we further have
\be
{\langle  {\cal O} (F'_+, F'_-)  \rangle}_{\tau +1, \alpha, \eta - \alpha} = {\langle  {\cal O} (F'_+, F'_-)  \rangle}_{\tau, \alpha, \eta},
\ee
and when $M$ is non-spin, we further have
\be
{\langle  {\cal O} (F'_+, F'_-)  \rangle}_{\tau +2, \alpha, \eta - 2\alpha} = {\langle  {\cal O} (F'_+, F'_-)  \rangle}_{\tau, \alpha, \eta},
\ee
all up to a $\tau$-independent $c$-number. In other words, the correlation function ${\langle  {\cal O} (F'_+, F'_-)  \rangle}_{\tau, \alpha, \eta}$ transforms like a modular form of $SL(2, \mathbb Z)$ (or $\Gamma_0(2)$) with weight $(a + {{\chi + \sigma} \over 4}, b + {{\chi - \sigma} \over 4})$, when $M$ is spin (or non-spin). Compare this with the weight $({{\chi + \sigma} \over 4}, {{\chi - \sigma} \over 4})$ of the partition function; hence, in general, we see that the partition function and correlation functions of the (non-singular) gauge-invariant local operators $F'_\pm$ all transform like modular forms but with different modular weights.

\vspace{1.0cm}
\hspace{-1.0cm}{\large \bf Acknowledgements:}\\
\vspace{-0.5cm}

This work is supported by the California Institute of Technology and the NUS-Overseas Postdoctoral Fellowship.

\vspace{0.0cm}


\end{document}